\documentclass[a4paper,fleqn]{cas-dc}

\usepackage[authoryear]{natbib}

\usepackage{hyperref}

\def\tsc#1{\csdef{#1}{\textsc{\lowercase{#1}}\xspace}}
\tsc{WGM}
\tsc{QE}

\begin{document}
\let\WriteBookmarks\relax
\def\floatpagepagefraction{1}
\def\textpagefraction{.001}
\let\printorcid\relax 

\shorttitle{Scale-Aware Cascaded Generative Mapping framework}

\shortauthors{Chenxing Sun et al.}

\title[mode = title]{Bridging Scales in Map Generation: A scale-aware cascaded generative mapping framework for seamless and consistent multi-scale cartographic representation}

\author[1]{Chenxing Sun}[style=chinese, orcid=0000-0001-7108-2988]
\credit{Conceptualization of this study, Methodology, Software}

\author[2]{Yongyang Xu}[style=chinese, orcid=0000-0001-7421-4915]

\author[3]{Xuwei Xu}[style=chinese]

\author[2]{Jing Bai}[style=chinese]

\author[2]{Xixi Fan}[style=chinese]

\author[4,5,6]{Xiechun Lu}[style=chinese, orcid=0000-0001-8450-2195]

\author[1,2,5,6]{Zhanlong Chen}[style=chinese, orcid=0000-0001-6373-3162]
\cormark[1]
\ead{chenzl@cug.edu.cn}

\address[1]{Key Laboratory of Geological Survey and Evaluation of Ministry of Education, China University of Geosciences, Wuhan 430074, China}
\address[2]{School of Computer Science, China University of Geosciences, Wuhan 430074, China}
\address[3]{School of Geographical and Information Engineering, China University of Geosciences, Wuhan 430074, China}
\address[4]{College Of Computer and lnformation Technology, China Three Gorges University, Yichang 443002, China}
\address[5]{Engineering Research Center of Natural Resource Information Management and Digital Twin Engineering Software, Ministry of Education, Wuhan 430074, China}
\address[6]{Hubei Key Laboratory of Intelligent Geo-Information Processing, China University of Geosciences, Wuhan 430074, China}

\cortext[1]{Corresponding author}

\begin{abstract}
    Multi-scale tile maps are essential for geographic information services, serving as fundamental outcomes of surveying and cartographic workflows. While existing image generation networks can produce map-like outputs from remote sensing imagery, their emphasis on replicating texture rather than preserving geospatial features limits cartographic validity. Current approaches face two fundamental challenges: inadequate integration of cartographic generalization principles with dynamic multi-scale generation and spatial discontinuities arising from tile-wise generation. To address these limitations, we propose a scale-aware cartographic generation framework (SCGM) that leverages conditional guided diffusion and a multi-scale cascade architecture. The framework introduces three key innovations: a scale modality encoding mechanism to formalize map generalization relationships, a scale-driven conditional encoder for robust feature fusion, and a cascade reference mechanism ensuring cross-scale visual consistency. By hierarchically constraining large-scale map synthesis with small-scale structural priors, SCGM effectively mitigates edge artifacts while maintaining geographic fidelity. Comprehensive evaluations on cartographic benchmarks confirm the framework's ability to generate seamless multi-scale tile maps with enhanced spatial coherence and generalization-aware representation, demonstrating significant potential for emergency mapping and automated cartography applications.
\end{abstract}




\maketitle



\section{Introduction}

\begin{figure}[htp]
	\centering
	\includegraphics[width=1.0\linewidth]{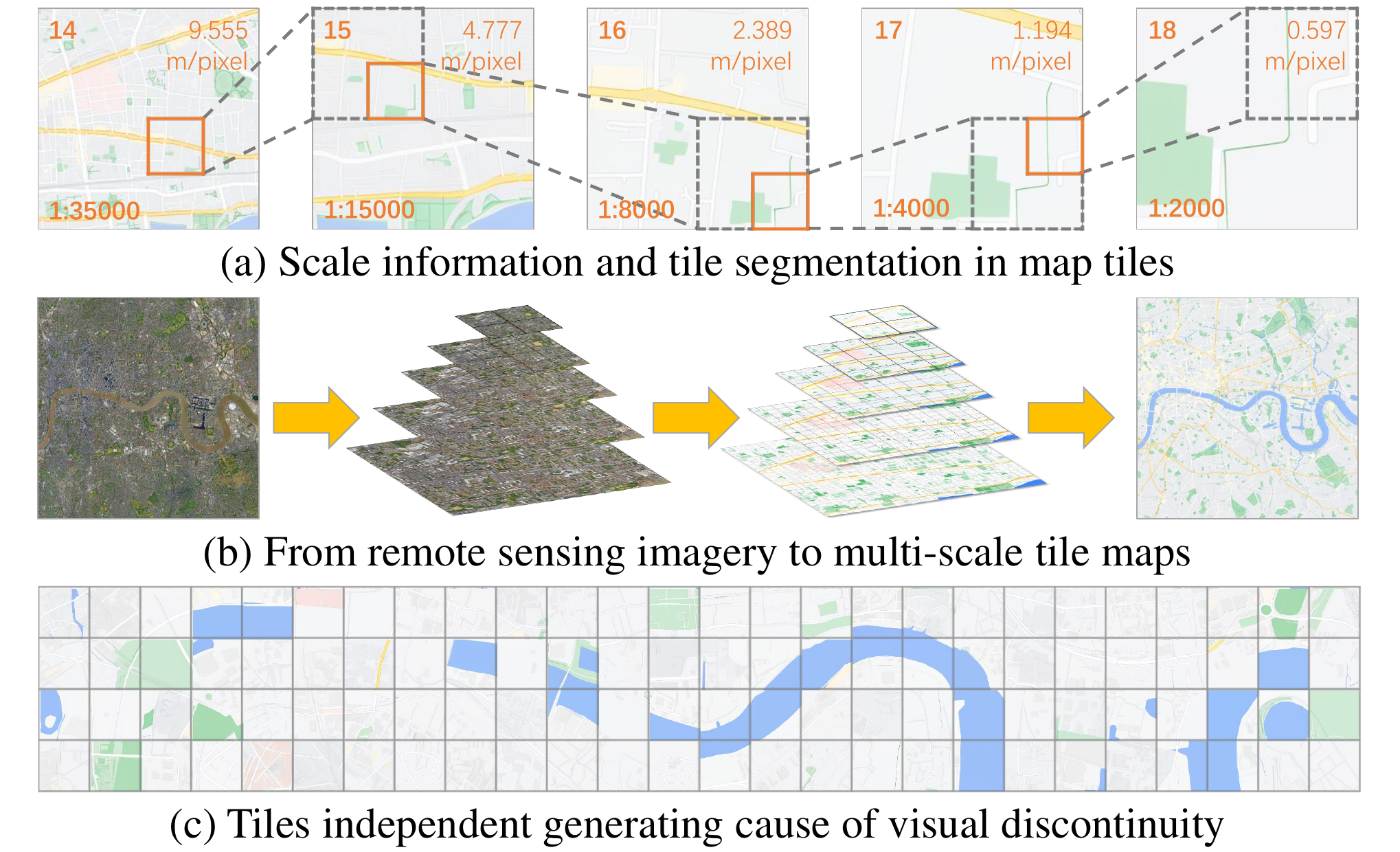}
    \caption{The generation of multi-scale maps from remote-sensing images can leverage scale information and tile segmentation in multi-scale tile map samples.}
	\label{fig:introduce-idea}
\end{figure}

Multi-scale tile maps, valued for their accessibility and readability, are integral to modern society, with extensive applications in emergency response, disaster management, and urban planning \citep{skidmore1997use,ezequiel2014uav}. The proliferation of online map services employing multi-scale tile maps, such as Google Maps, Bing Maps, and OpenStreetMap, has made it easy for individuals to access and utilize these resources \citep{stefanakis2017web}. Although traditional mapping methods generally depend on field surveys, modern map providers now frequently employ remote sensing and mapping vehicles for data collection. However, these cartographic techniques still heavily rely on the manual acquisition of vector data, cartographic generalization and manual selection, hindering the delivery of real-time map services \citep{haunold1993keystroke,park2011hybrid}. In practical scenarios, such as mapping remote areas or conducting emergency rescue operations, the rapid generation of tiled maps and real-time updates of map services is particularly crucial. Advancements in remote-sensing technology now allow satellites, drones, and aircraft to capture high-resolution images rich in geographical information rapidly \citep{shen2022high,liu2022raanet,zhao2024rsmamba}. This development makes it feasible to create multi-scale map tiles from real-time remote sensing data (Figure\ref{fig:introduce-idea} (b)). Recent advancements in image generation offer a more efficient alternative to traditional cartographic workflows\citep{goodfellow2014generative,ganguli2019geogan}. Such an approach significantly accelerates map production and ensures timely updates, particularly in inaccessible or hazardous areas for ground-based surveys.

Current map generation model, exemplified by MapGAN \citep{li2020mapgan}, SMAPGAN \citep{chen2021smapgan}, and CreativeGAN \citep{fu2022translation}, utilize image translation \citep{isola2017imagetoimage,wang2018highresolution} methods to achieve map creation from remote sensing imagery. However, these approaches remain constrained to fixed-scale, single-layer map generation paradigms. Specifically, they exhibit insensitivity to hierarchical scale information inherent in tile-based maps and lack the capacity for multi-scale tile map production. Moreover, Multi-level map generate method can generate multi-scale maps using map-level \citep{liu2021cscgan,fu2023levelaware}, but fixed level encoding fails to provide a detailed understanding of map scales and overlooks cartographic knowledge associated with scale modality information. The current flaws of applying generation method in multi-scale cartography from remote sensing images lay in three aspects:

1) Single-scale models fail to generate coherent multi-scale maps essential for hierarchical decision-making. Leveraging cartographic expertise from extensive tile map samples (Fig. \ref{fig:introduce-idea}(a)) to improve scale accuracy remains challenging. Current methods rigidly link map levels to outputs, neglecting the integration of scale-specific characteristics like spatial resolution, map scale, and visual features. This oversight limits models’ ability to utilize multi-scale cartographic knowledge, causing inaccurate scale representation and reduced practicality of generated maps.

2) Tile-by-tile generation frequently introduces visual inconsistencies along tile edges after stitching (Fig. \ref{fig:introduce-idea}(c)), which compromises both spatial readability and practical usability. These artifacts are particularly problematic in emergency mapping scenarios where efficiency and accuracy are critical. Additionally, stylistic inconsistencies across tiles negatively impact the map's overall coherence and aesthetic quality. Furthermore, the absence of comprehensive geographical semantic information in tile samples hinders the effectiveness of generative model training.

3) Single-scale generative models are inadequate for dynamically and accurately representing map features in practical mapping scenarios, failing to seamlessly integrate cartographic features from macro to micro perspectives. Moreover, reliance on isolated geographical semantic extraction modules restricts spatial context understanding, while semantic segmentation samples risk obscuring detailed map information. These limitations impair cross-modal spatial comprehension and reduce the generative model's expressive capabilities.

To address these challenges, we propose a novel framework, SCGM, for generating multi-scale tile maps from remote sensing imagery. By leveraging conditional diffusion models and a multi-scale cascade generation approach, SCGM effectively integrates scale information and cascading references to capture the intricate relationships between map scales and geographic features. Unlike previous single-stage methods, SCGM recursively generates multi-scale tile maps while incorporating controllable parameters, such as map scale, level, and resolution, to facilitate the production of larger-scale tiles. This unified framework enables the parallel generation of spatially consistent multi-scale tile maps for large-format scenes, as demonstrated on the CSCMG dataset — a comprehensive, real-world, multi-scale remote sensing mapping dataset that utilizes associated scale information.

In summary, our main contributions are as follows:

1) We have developed a scale-modality encoding mechanism to formalize map generalization relationships. This approach dynamically achieves data-driven embedding of multi-scale cartographic knowledge priors, aligns cartographic generalization with geographic semantics, and effectively integrates the expertise of cartographic synthesis specialists into the generative model.

2) We eliminate edge discontinuities while maintaining semantic coherence across scales by employing smaller scale map tiles as spatial constraints in a cascaded generation strategy. It enables boundless map generation for uncharted regions, where conventional methods produce fragmented outputs unsuitable for emergency coordination.

3) Our dual-branch scale-adaptation network extracts complementary features from remote sensing imagery guided by cascaded references. The architecture guarantees cartographic fidelity across various scenarios by adaptively weighting local textures and global structures, providing a rich representation of intricate geographical features in the resultant map.

\section{Related work}

\subsection{Image-to-image generation methods}

Image-to-image (I2I) translation is crucial in cross-domain visual representation learning, particularly in converting remote sensing data into structured mapping outputs. Early methods predominantly employed generative adversarial networks (GANs) \citep{goodfellow2014generative,mirza2014conditional}, with pioneering works such as Pix2Pix \citep{isola2017imagetoimage}, Pix2PixHD \citep{wang2018highresolution} and CycleGAN \citep{zhu2017unpaired} establishing both supervised and unsupervised frameworks for cross-domain mapping. Subsequent innovations, including StarGAN \citep{choi2018stargan}, expanded GAN-based architectures to accommodate multi-domain translation tasks through unified network designs. Although these approaches achieved pixel-level consistency and semantic preservation, challenges continued to arise in modeling high-dimensional feature distributions and synthesizing photorealistic details \citep{pang2022imagetoimage}.

Recent diffusion-based generators mitigate key limitations of GANs — specifically, mode collapse and adversarial instability — by employing likelihood-based training, which enhances sample diversity and convergence \citep{sohl-dickstein2015deep}. DDPMs \citep{ho2020denoising} pioneered this paradigm through a two-phase process: gradually corrupting data with Gaussian noise (forward) and learning to reverse it (reverse). Subsequent work improved training efficiency — Nichol et al. optimized noise variance estimation \citep{nichol2021improved}, while DDIM \citep{song2022denoising} reformulated diffusion as non-Markovian, enabling faster deterministic sampling without quality loss.

Conditional diffusion models extend this framework by incorporating external guidance (e.g., class labels, text, images) for controlled synthesis \citep{ho2022cascaded, saharia2022palette, rombach2022highresolution}. However, pixel-space diffusion remains computationally prohibitive for high-resolution outputs. Two dominant solutions address this: cascaded generation \citep{saharia2022photorealistic}, which progressively upscales low-resolution intermediates, and latent diffusion \citep{rombach2022highresolution}, which operates in compressed feature spaces. Both strategies balance fidelity and efficiency, enabling scalable synthesis of complex imagery.

Consequently, these approaches frequently fail to adequately extract and represent the intricate features of objects in remote sensing imagery, resulting in maps of subpar quality compared to those produced by methods specifically tailored for this domain.

\subsection{Map generation from RS image}

Researchers have employed generative techniques like GANs to enable image-to-map translation by learning pixel-level mappings from remote sensing data \citep{isola2017imagetoimage}. While methods like GeoGAN \citep{ganguli2019geogan}, SMAPGAN \citep{chen2021smapgan}, and CreativeGAN \citep{fu2022translation} enhance feature fidelity through conditional architectures and semantic constraints, they predominantly focus on single-scale outputs, failing to capture hierarchical geographic relationships essential for multi-scale mapping.

Efforts to address multi-scale challenges remain fragmented. Chen et al. \citep{chen2022generating} proposed hierarchical generation using separate models per scale, but error propagation from coarse-to-fine levels limits accuracy. CscGAN \citep{liu2021cscgan} unified multi-level generation via a single model but required auxiliary road morphology inputs, restricting scalability. The SOTA method LACG \citep{fu2023levelaware} introduced level-aware fusion but suffers from tile-wise inconsistencies due to independent tile generation, undermining spatial coherence in composite maps.

In conclusion, existing methods either neglect cross-scale semantic dependencies or fail to ensure visual continuity across tiles, hindering their utility in emergency mapping and large-format applications. Our work addresses these challenges through a cascaded conditional diffusion framework, enabling end-to-end synthesis of seamless, geographically consistent multi-scale tile maps from unified remote sensing inputs.

\section{Method}

\subsection{Overall framework}

\begin{figure*}[htp]
	\centering
	\includegraphics[width=1.0\linewidth]{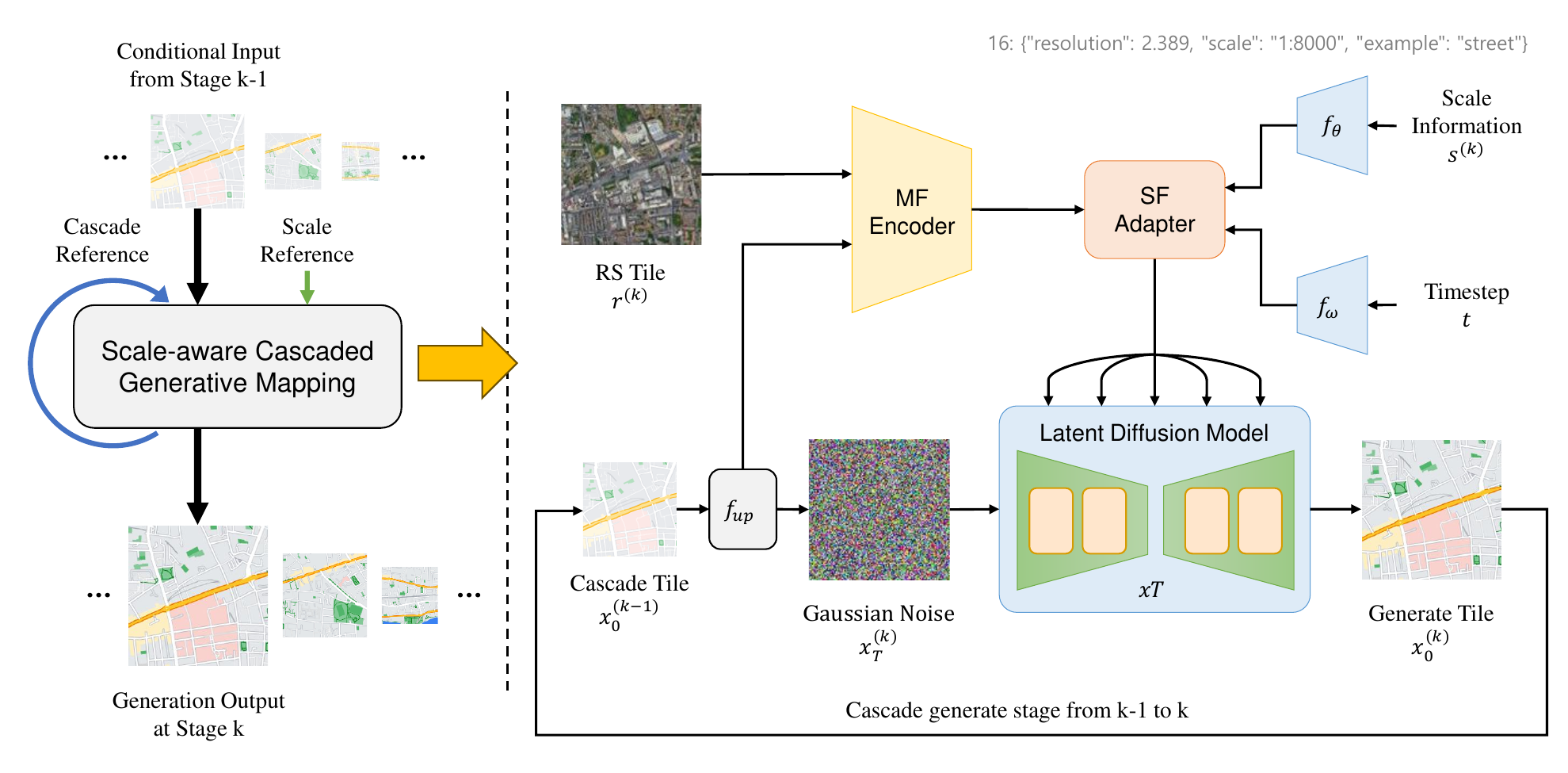}
	\caption{
		Overview of the SCGM framework: a hierarchical, self-cascading pipeline that progressively synthesizes multi-scale tile maps by refining smaller-scale outputs under explicit scale conditioning.
	}
	\label{fig:method-framework}
\end{figure*}

We propose SCGM, a self-cascading generative mapping framework for multi-scale tile map synthesis, following the architecture of LDM \citep{rombach2022highresolution}. As shown in Figure \ref{fig:method-framework}, SCGM employs a unified generative model to recursively construct large-scale maps from smaller-scale representations. The framework comprises three core modules: ScaleEncoder, which embeds scale-related spatial context; MFEncoder, which fuses semantic features from remote sensing imagery and cascading references; and SFAdapter aligns multi-scale features with the denoising trajectory.

The generation proceeds in stages, with each step incorporating three inputs: remote sensing images for geographic content, lower-scale maps as spatial priors, and scale information guiding generalization. Unlike resolution-focused models \citep{ho2022cascaded,tian2024visual}, SCGM emphasizes scale-specific semantics, enabling hierarchical abstraction consistent with cartographic conventions. This structured cascade supports coherent multi-scale synthesis, with further details on feature integration and scale adaptation presented later.

At generation stage \(k\), let \(x_0^{(k)}\) denote the target map and \(s^{(k)}\) the associated scale information, including map layer \(s_z^{(k)}\), spatial resolution \(s_r^{(k)}\), scale ratio \(s_s^{(k)}\), and geospatial encoding \(s_f^{(k)}\). To generate the map at stage \(k+1\), the model conditions on \(x_0^{(k)}\) and \(s^{(k)}\). Assuming the model advances the tile level by \(n\) steps per stage (e.g., \(n = 1\)), an input tile of size \(H \times W\) pixels at level \(s_z^{(k)}\) results in \(4^n\) output tiles at level \(s_z^{(k)} + n\), collectively spanning \(2^n H \times 2^n W\) pixels over the same geographic region.

This process can be iterated \(m\) times, enabling parallel generation of increasingly detailed maps across multiple scales. The complete multi-scale output is expressed as:
\begin{eqnarray}
    \mathcal{X} = \{ x_0^{(k)}, x_0^{(k+1)}, \dots, x_0^{(k+m)} \}
    \end{eqnarray}
where \(x_0^{(k+m)}\) corresponds to level \(s_z^{(k)} + m \cdot n\) and covers an area of size \(H \cdot 2^{nm} \times W \cdot 2^{nm}\) pixels.

\subsection{Conditional denoising map generation}

Map generation relies solely on remote sensing imagery tiles as inputs to produce corresponding spatial scene map tiles. This conditional denoising process can be mathematically expressed as follows:
\begin{eqnarray}
	y=\mathop{\arg\min}_{y}-\log\left(p\left(y\mid x\right)\right)
\end{eqnarray}
where \(x\) denotes the remote sensing imagery tiles, while \(y\) represents the generated map tiles. The term \(p(\cdot \mid x)\) refers to the conditional probability density function of the target map given the remote sensing imagery input, typically modeled as either a standard Gaussian distribution (\(L_2\) loss) or a Laplacian distribution (\(L_1\) loss), with the true target map serving as its mean. This approach, which relies exclusively on remote sensing imagery tiles, theoretically learns the many-to-one mapping between remote sensing tiles and the corresponding map tiles. However, when applied to multi-scale map generation, these methods often need to be revised, struggling to extract precise semantic information about geo-objects across multiple scales of remote sensing imagery.

To address these limitations, drawing inspiration from advancements in decoupled feature learning, we propose the following revised modeling approach:
\begin{eqnarray}
	y=\mathop{\arg\min}_{y}-\log\left(p\left(y\mid x,con,rsm\right)\right)
	\label{equ:SCGM_model}
\end{eqnarray}
where \(con\) and \(rsm\) represent scale information and cascading reference conditions. Both serve as pivotal components for encoding cartographic features. In the context of this methodology, the terms are defined as follows:
\begin{itemize}
    \item Scale Information: The spatial resolution, map level, and feature types associated with the specified scale of the generated map.
    \item Cascade Reference: The smaller-scale map tiles produced in the previous stage serve as geographic feature cascading constraints.
\end{itemize}

\begin{figure*}[tp]
	\centering
	\includegraphics[width=\linewidth]{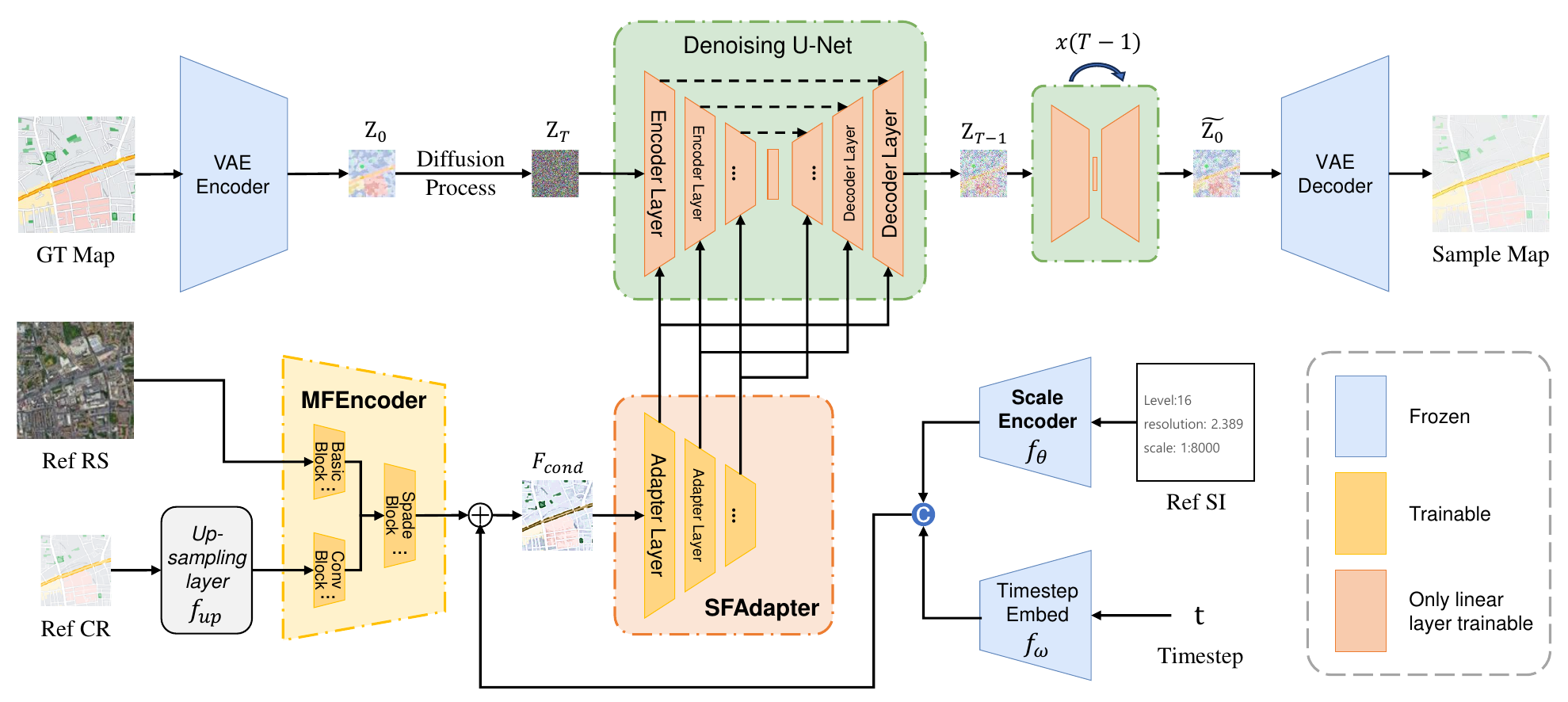}
	\caption{The SCGM architecture leverages a VAE to transition the diffusion and reverse processes from pixel space to latent space. During training, the latent representation of the target map, \(z_0\), undergoes a progressive transformation into \(z_t\) through the diffusion process, followed by denoising achieved via a U-Net network. Two bespoke modules are introduced to further refine the denoising process: MFEncoder and SFAdapter. The MFEncoder integrates information derived from remote sensing imagery and cascading references to construct the conditional feature, \(F_{\mathrm{cond}}\). Expanding on \(F_{\mathrm{cond}}\), the SFAdapter produces multi-scale features that are subsequently merged with the outputs of corresponding U-Net layers through element-wise addition.}
	\label{fig:method-model}
\end{figure*}

\subsection{Scale information guidance}

At each generation stage, previously generated smaller-scale tiles, along with scale information embeddings, serve as conditional inputs. Since the cascading reference map \(x_0^{(k)}\) differs in resolution from the current input \(x_t^{(k+1)}\), we adapt the architecture following principles from \citep{saharia2022image, li2022srdiff}. As shown in Figure \ref{fig:method-model}, the encoder \(E_{lr}\) first extracts features from \(x_0^{(k)}\), which are then upsampled and transformed via a module \(f_{up}\) comprising convolutional and interpolation layers to match the spatial dimensions of \(x_t^{(k+1)}\). The resulting features are concatenated with \(x_t^{(k+1)}\) along the channel dimension for fusion.

At each generation stage, the previously generated small-scale tiles, along with time step and scale information embeddings are utilized as conditional input variables. For the smaller-scale tile input — cascading reference map \(x_0^{(k)}\) — its dimensions do not align with \(x_t^{(k+1)}\). To address this, and inspired by \citep{saharia2022image} and \citep{li2022srdiff}, we redesigned the network architecture. First, as illustrated in Figure \ref{fig:method-framework}, the encoder \(E_{lr}\) is applied to encode the smaller-scale map tiles \(x_0^{(k)}\). Then, a series of upsampling and convolutional layers \(f_{up}\) are employed to align the feature map dimensions with \(x_t^{(k+1)}\). Finally, the features are fused by concatenating \(x_t^{(k+1)}\) and the feature maps along the channel dimension:
\begin{eqnarray}
	\tilde{x}_t^{(k+1)}=\text{cat}[x_t^{(k+1)},f_{up}(E_{lr}(x_0^{(k)}))]
\end{eqnarray}
where \(\text{cat}(\cdot)\) represents the concatenation operation along the channel dimension.

For the scale information \(s\) in text modality data, we perform cross-modal encoding using pre-trained CLIP model \citep{radford2021learning,cherti2023reproducible}, with the ScaleEncoder denoted as \(f_\theta\) (Figure \ref{fig:method-model}). This approach allows us to obtain the embedding vector of scale information \(e_s^{(k)} = f_\theta(s)\) at each generation stage, capturing the cross-modal representation of scale information corresponding to the spatial context of the generated map. This embedding subsequently guides the map generation process through the cross-attention mechanism \citep{vaswani2017attention} during denoising:
\begin{eqnarray}
	e_s^{(k)}=f_\theta(s_z^{(k)}+s_r^{(k)}+s_s^{(k)}+s_f^{(k)})
\end{eqnarray}

Through this process, the scale information embedding \(e_s^{(k)} \in \mathbb{R}^D\) is obtained at the \(k\)-th generation stage. The time step variable \(t \in \{1, \dots, T\}\) is then encoded via a frequency encoding transformation \(f_\omega\):
\begin{eqnarray}
	e_t=f_\omega(t)
\end{eqnarray}
where \(\theta_t\) is a learnable parameter.

Finally, \(e_s^{(k)}\) and \(e_t\) are combined to generate the final conditional embedding vector for the \(k\)-th generation stage and the \(t\)-th denoising step:
\begin{eqnarray}
	e_t^{(k)} = e_s^{(k)} + e_t
\end{eqnarray}

Using the conditional embedding vector \(e_t^{(k)}\) and the cascade map feature \(\tilde{x}_t^{(k)}\), we define the conditional variable for the \(k\)-th generation stage and \(t\)-th denoising step as:
\begin{eqnarray}
	c_t^{(k)} = \{e_t^{(k)}, \tilde{x}_t^{(k)}\}
\end{eqnarray}

This condition \(c = c_t^{(k)}\) is then used as input to model the following conditional probability density function \(p_\theta\left(y \mid x, c\right)\):
\begin{align}
    p_\theta(x_{0:T}|c_t^{(k)})&=p(x_T)\prod_{t=1}^{T}{p_\theta(x_{t-1}|x_t, c_t^{(k)}}) \\
    p_\theta(x_{t-1}|{x_t},c_t^{(k)})&=\mathcal{N}(x_{t-1};\mu_\theta(x_t,t,c_t^{(k)}),\sigma_t^2{I})
\end{align}

Based on Equation \eqref{equ:SCGM_model}, while keeping the forward process unchanged, we construct a conditional denoising model:
\begin{eqnarray}
	y=\mathop{\arg\min}_{y}-\log\left(p_\theta(x|c)\right)
\end{eqnarray}

Through this design, we can ultimately generate map tiles of varying scales in a self-cascading manner, encoded with information on temporal steps, and scale information.

\begin{figure}[tp]
	\centering
	\includegraphics[width=\linewidth]{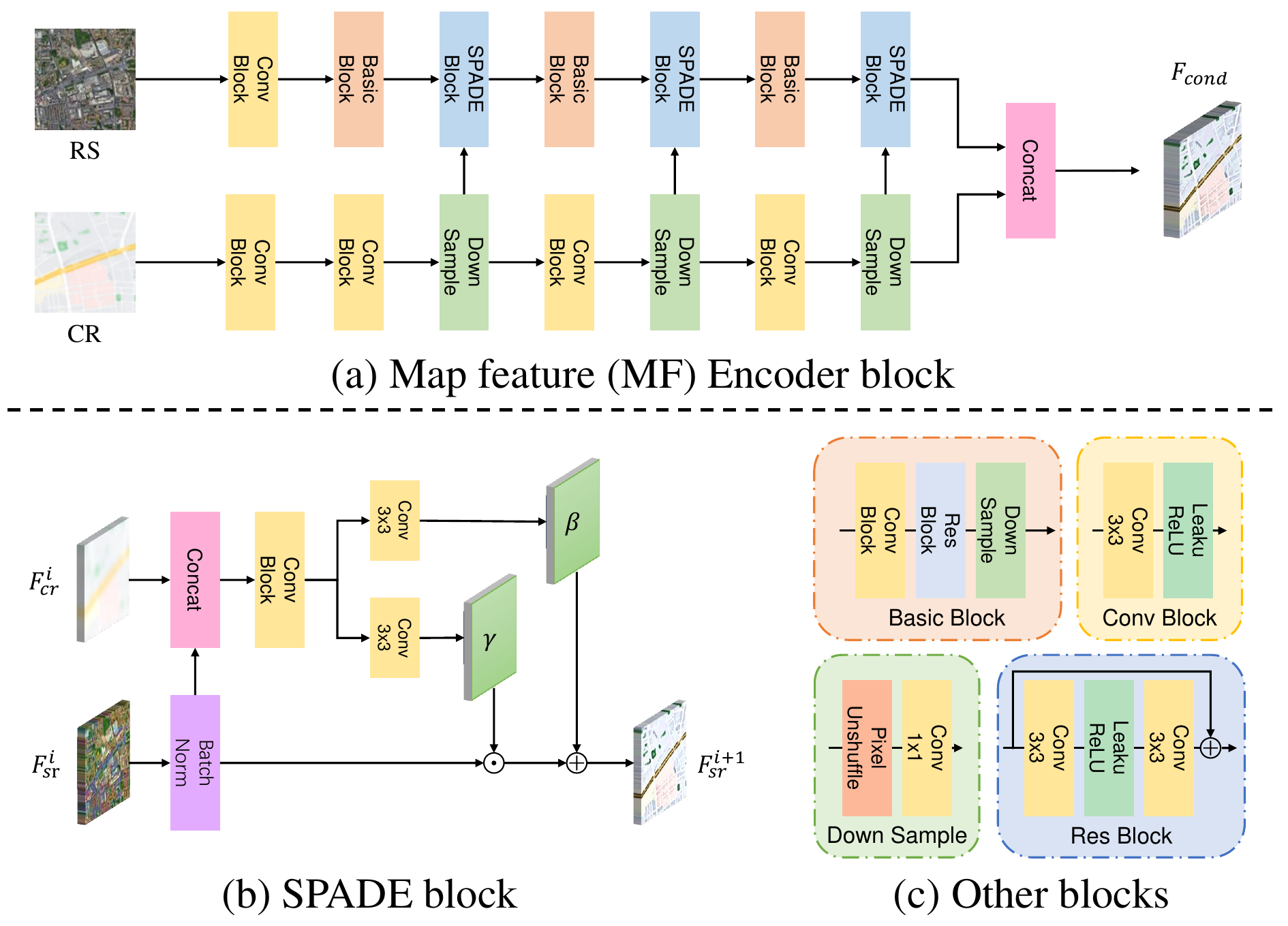}
	\caption{Detailed structure of the MFEncoder, SPADE module and basic blocks.}
	\label{fig:method-blocks}
\end{figure}

\subsection{Map feature encoder}

To enable effective integration of semantic cues from cascading reference tiles into remote sensing image tiles, we propose a multi-branch map feature encoder MFEncoder, as shown in Figure \ref{fig:method-model}. This encoder incorporates multi-scale semantic information from the reference tiles into the remote sensing features. As illustrated in Figure \ref{fig:method-blocks}(a), one branch extracts hierarchical features from the reference tiles using convolutional and downsampling blocks, while the other branch processes remote sensing image tiles through a series of basic blocks.

We employ the SPADE block \citep{park2019semantic} at each stage to facilitate feature fusion across scales. The SPADE block utilizes remote sensing features—characterized by coarse textures and structural details—as its primary input in our design. As illustrated in Figure \ref{fig:method-blocks}(b), features from both remote sensing and reference tiles are jointly leveraged to compute the scale (\(\gamma\)) and bias (\(\beta\)) parameters. This design enables more adaptive and semantically aligned fusion, enhancing the fidelity of content-aware feature integration \citep{wang2025semantic}. The structure of each module, encompassing basic, convolutional, and downsampling blocks, is depicted in Figure \ref{fig:method-blocks}(c).

\subsection{Scale feature adapter}

The SFAdapter module (Figure \ref{fig:method-model}) transforms the dimensions of scale-mapped features received from the encoder to generate scale-specific feature representations compatible with the latent space organization of the pre-trained diffusion model. It synthesizes the final comprehensive conditional features \(F_{\mathrm{cond}}\) by integrating texture and structural patterns extracted from image patches by the MFEncoder with contextual attributes obtained from the cascading reference blocks. The SFAdapter demonstrates exceptional capability in fusing multimodal features.

Specifically, the architecture of the SFAdapter module is straightforward, as detailed in Figure \ref{fig:method-blocks}(c), employing sequential processing stages through convolutional operations, residual connections, and progressive downsampling. The module systematically expands feature channel capacity through successive transformations while contracting spatial resolution, producing three hierarchical conditional features at varying scales. These multi-scale representations are subsequently merged with corresponding hierarchical features in the U-Net decoder through channel-wise concatenation and additive fusion, thereby enabling spatial-adaptive modulation of the diffusion model's generation pathway.

\subsection{Loss function} \label{sec:loss}

The trainable parameters of the SCGM are determined by the following loss function:
\begin{eqnarray}
    L=\mathrm{\mathbb{E}}_{\left(z_{0},\epsilon,rs,cr,t\right)}\parallel \epsilon-\epsilon_{\theta}\left(z_{t},t,rs,cr\right) \parallel ^{2}
\end{eqnarray}
this eqnarray, \(rs\) and \(cr\) represent the input remote sensing image tiles and cascading reference tiles. Here, \(\epsilon\) denotes randomly sampled Gaussian noise, while \(t\) signifies the integer timestamp sampled from the interval \([0, T]\). In this context, \(\epsilon_{\theta}\) refers to the denoising U-Net model.

\begin{figure*}[hbp]
    \centering
    \includegraphics[width=\linewidth]{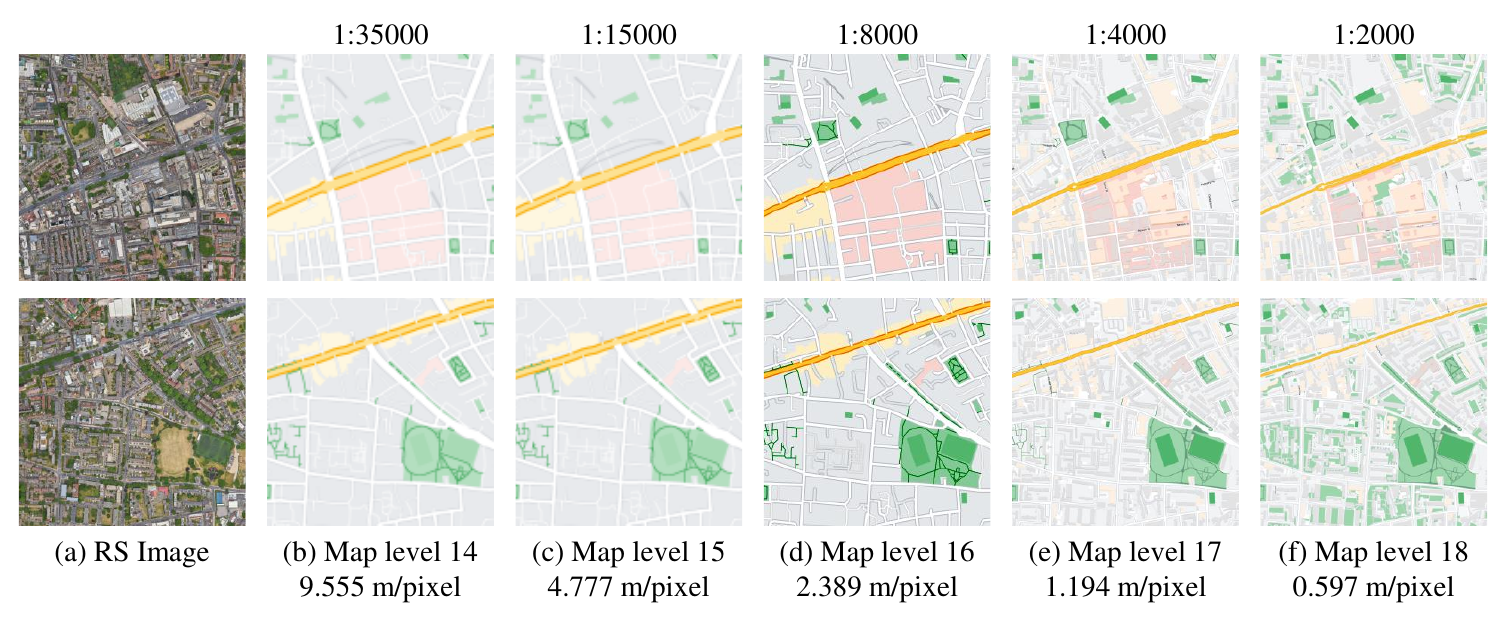}
    \caption{Examples of RS-Map tile sample pairs from scales 1:35,000 to 1:2,000 in the CSCMG dataset are provided, covers a broad range of cross-scale scenarios.}
    \label{fig:dataset-example}
\end{figure*}

\subsection{Evaluation metrics} \label{sec:metrics}

We adhere to established practices in generative mapping \citep{fu2023levelaware,fu2022translation} by employing quality evaluation metrics to comprehensively assess the performance of map generation methods. The realism of the generated maps is quantified using the PSNR and SSIM, calculated on the Y channel within the YCbCr color space \citep{wang2004image}. Furthermore, the FID is utilized to evaluate the distributional divergence between real and generated maps \citep{heusel2017gans}.

While semantic segmentation metrics (e.g., mIoU) could offer deeper insights, our method focuses on large-scale, paired image-to-image translation without pixel-level semantic labels. To address the limitations of pixel or distribution-based metrics in assessing semantic-level cartographic features, we employ the map feature perception metric MFP \citep{sun2025map}. This metric complements traditional evaluations by explicitly measuring global semantic alignment and spatial consistency between synthesized and reference maps. The MFP leverages a self-supervised Vision Transformer (ViT) pretrained with DINO \citep{caron2021emerging} to extract deep features, comprising two components:

Global feature evaluation captures holistic cartographic attributes (e.g., scene categories, element distributions) using the [CLS] token embeddings from the final ViT layer:
\begin{eqnarray}
    \mathcal{L}_G = 1 - \text{MSE}(\text{CLS}_o, \text{CLS}_t)
\end{eqnarray}
where $\text{CLS}_o$ and $\text{CLS}_t$ denote embeddings for generated and target maps, respectively.

Spatial similarity evaluation quantifies local geometric relationships via self-attention keys from the final ViT layer:
\begin{eqnarray}
    \mathcal{L}_S = \left\| S(I_o) - S(I_t) \right\|_F
\end{eqnarray}
where $S(I) = \text{cos\_sim}(K^L(I))$ computes the cosine similarity matrix of patch-wise semantic correlations.

The integrated MFP is defined as:
\begin{eqnarray}
    \text{MFP} = \lambda_1 \mathcal{L}_G + \lambda_2 \mathcal{L}_S
\end{eqnarray}
with $\lambda_1=10$ and $\lambda_2=1$ empirically determined. Lower MFP values indicate superior semantic fidelity and spatial coherence. A detailed explanation of the interpretability and spatial consistency of the MFP metric is provided in \citep{sun2025map}.

\section{Experiment}

\subsection{Experimental setup}

\subsubsection{Datasets}\label{sec:Datasets}

To address limitations in existing multi-scale map datasets, we present the cross-scale cascade map generation dataset CSCMG, comprising real-world remote sensing images and tile maps from Glasgow and London, UK. CSCMG pairs multi-scale tile maps with corresponding imagery, sampled from non-overlapping Glasgow regions for training/testing and continuous London areas for seamless generation evaluation. Each pair includes scale information (map scale, level, resolution), enabling cross-scale validation. Representative examples are shown in Figure \ref{fig:dataset-example} and the detailed information shown in Table \ref{tab:cmmgd_dataset}.

\begin{table*}[ht]
    \caption{The detailed information of the CSCMG dataset, including the quantity of RS-Map tile sample pairs across various scales, as well as scale information (map scale, level, resolution, and feature types).}
	\label{tab:cmmgd_dataset}
    \centering
    \begin{tabular*}{\textwidth}{@{\extracolsep{\fill}}cccccc}
    \toprule
    \multirow{2}{*}{ Level} & \multicolumn{2}{c}{Tile pair number} & \multirow{2}{*}{Scale} & \multirow{2}{*}{\begin{tabular}[c]{@{}c@{}}Resolution\\ (m/pixel)\end{tabular}} & \multirow{2}{*}{Feature types} \\
        & Training set & Testing set &  &  &  \\ \midrule
     14 & 4870 & 280 & 1:35,000 & 9.555 & village, or suburb \\
     15 & 15106 & 292 & 1:15,000 & 4.777 & small road \\
     16 & 30040 & 300 & 1:8,000 & 2.389 & street \\
     17 & 40555 & 300 & 1:4,000 & 1.194 & block, park, addresses \\
     18 & 45000 & 300 & 1:2,000 & 0.597 & some buildings, trees \\
     All & 135571 & 1471 & - & - & - \\ \bottomrule
    \end{tabular*}
\end{table*}

\begin{figure}[ht]
    \centering
    \includegraphics[width=\linewidth]{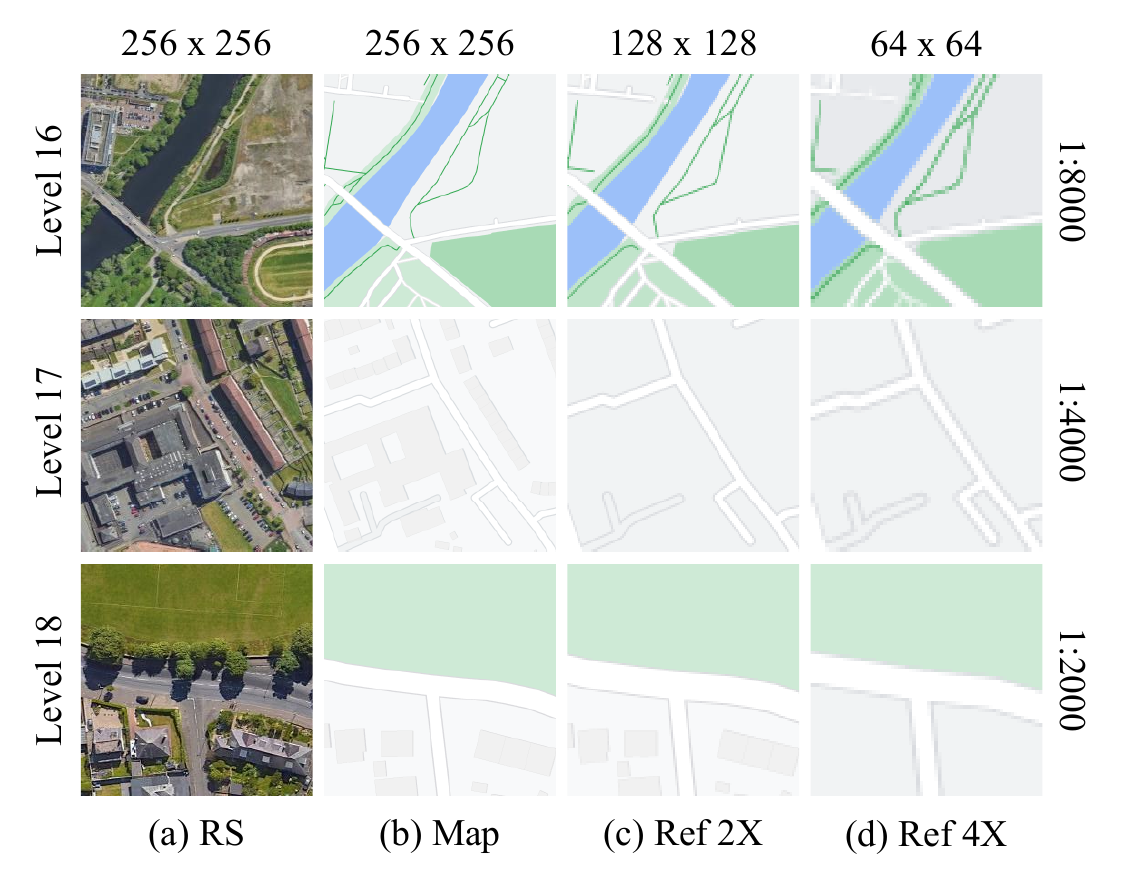}
    \caption{Examples of cascade references from scales of 1:8,000 to 1:2,000 in the CSCMG dataset. The dataset provides 2X or 4X cascade references, with image resolutions of 128 and 64.}
    \label{fig:dataset-cascade}
\end{figure}

The CSCMG dataset comprises 137,042 paired remote sensing and map tiles spanning scales from scales 1:35,000 to 1:2,000, sourced from Google Maps’ free tile service. These tiles are organized through quadtree division to hierarchically span spatial extents ranging from scales of 1:35,000 (the coarsest) to 1:2,000 (the finest), including 135,571 tile pairs for training and 1,471 for testing. As illustrated in Figure \ref{fig:dataset-cascade}, cascade references at 2X and 4X scales provide multi-scale constraints: reference tiles (64 or 128 resolution) are upscaled to 256×256 via bi-cubic resampling, ensuring uniform tile sizes while preserving spatial resolution relationships. This design verifies multi-scale cascading generation efficacy, particularly in hierarchical feature alignment and seamless map tile synthesis.

To evaluate our method's multi-scale map generation capabilities, we utilize the multi-scale map generation dataset MLMG constructed by the SOTA method LACG \citep{fu2023levelaware}. The MLMG comprises samples collected from cities in the US and China, reflecting significant differences in their respective images and maps. In China, roads are typically wider with lower distribution density, lacking a clear parallel-vertical relationship, whereas in the United States, roads are generally straight and densely packed, with the ground neatly segmented into blocks.

\subsubsection{Implementation details}
The proposed SCGM framework is developed on the Stable Diffusion pretrained model (SD-2.1-base) and implemented using PyTorch. All experiments are performed on dual NVIDIA GeForce RTX 3090 GPUs, utilizing cascaded map tiles and remote sensing imagery from scales 1:35,000 to 1:2,000 of the CSCMG dataset.

In the training process, the pretrained model is fine-tuned using the loss function described in Section \ref{sec:loss}. Optimization targets include the Map Feature Encoder (MFEncoder), Scale Feature Adapter (SFAdapter), and the linear layers of the U-Net. The model contains 152 million trainable parameters out of a total of 1.4 billion and adopts a hierarchical structure to reduce redundancy via cross-scale parameter sharing. Training is conducted for 200,000 steps with a batch size of 20, using the AdamW optimizer with a learning rate of \(5 \times 10^{-5}\), and converges within 480 GPU-hours on two RTX 3090 GPUs.

\begin{figure*}[htp]
    \centering
    \includegraphics[width=\linewidth]{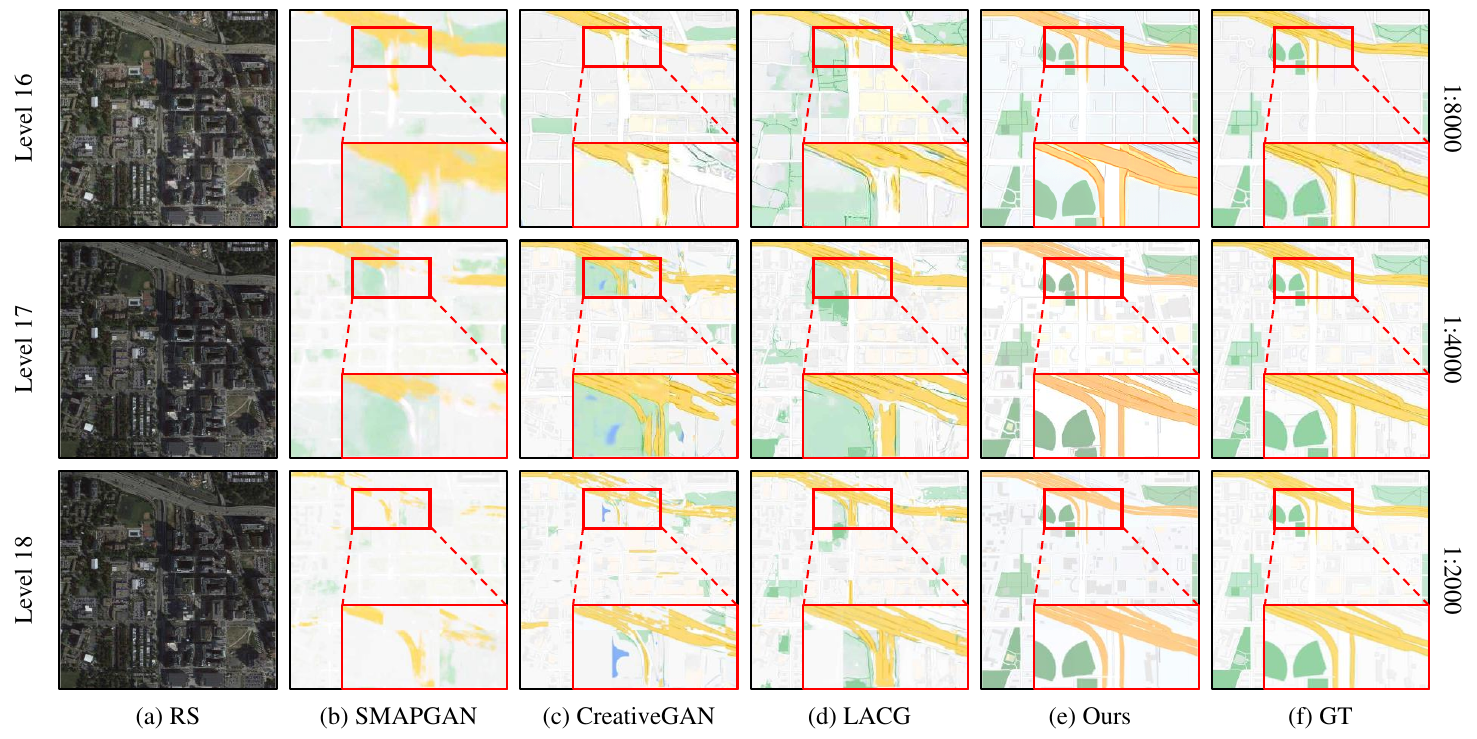}
    \caption{Generated results on the MLMG-US test set using SCGM and baseline methods, highlighting SCGM’s superior seamless tile synthesis and geographic detail preservation across scales.}
    \label{fig:experiment-qualitative}
\end{figure*}

\subsection{Experimental results}

To substantiate the efficacy of our technique, we compared our method against established image-to-image translation methods (Pix2pix \citep{isola2017imagetoimage}, Pix2pixHD \citep{wang2018highresolution}, CycleGAN \citep{zhu2017unpaired}, SPADE \citep{park2019semantic}, SelectionGAN \citep{tang2019multichannel}), map generation methods (SMAPGAN \citep{chen2021smapgan}, Creative GAN \citep{fu2022translation}), and the current SOTA method, LACG \citep{fu2023levelaware}.

\subsubsection{Quantitative evaluation}

SCGM consistently achieved optimal outcomes on average across MLMG datasets, as shown in Table \ref{tab:comparison}. Specifically, on the MLMG-CN dataset, the proposed method recorded an FID index of 103.99 and a PSNR index of 29.759, outperforming other methods. Similarly, the MLMG-US dataset achieved an FID index of 104.63 and a PSNR index of 28.012, demonstrating superior performance. These results indicate that our method surpasses others in these image quality evaluation metrics, effectively validating its efficacy. Notably, the FID results highlight a significant advantage, suggesting that the distribution difference between the maps generated by our method and the real maps is minimal, thus reflecting high-quality output.

Furthermore, we identified suboptimal results in the testing outcomes; for instance, the LCAG method exhibited a suboptimal performance in the FID metric, underscoring the substantial advantage of our method in terms of perceived quality compared to LCAG. In the context of the PSNR metric, our method outperformed the suboptimal SMAPGAN method, indicating higher similarity in the generated maps. However, it is important to note that some researchers have pointed out that a higher PSNR does not necessarily correlate with improved visual quality \citep{ledig2017photorealistic,pei2018does}. We opted to retain this metric as it provides a relative measure of the generated map quality across different methods and is a valuable reference in ablation studies.

\begin{table}[t]
    \caption{Quantitative comparison of disparate methodologies on the MLMG dataset, with bold entries indicating optimal results and the underlined entries denoting the suboptimal results.}
    \label{tab:comparison}
    \centering
    \begin{tabular}{lcccc}
        \toprule
        \multirow{2}{*}{Model} & \multicolumn{2}{c}{MLMG-CN} & \multicolumn{2}{c}{MLMG-US} \\
        &  FID$\downarrow$ & PSNR$\uparrow$  &  FID$\downarrow$ & PSNR$\uparrow$  \\
        \midrule
        Pix2Pix & 227.31 &  23.16  &  248.66 &  26.406 \\
        Pix2PixHD   & 195.48 &  23.044 &  188.22 &  25.297 \\
        CycleGAN    & 213.99 &  22.515 &  153  & 24.85  \\
        SPADE  & 332.78 &  23.166 &  304.46 &  25.948 \\
        SelectionGAN  &  261.07 &  23.169 &  260.18 &  25.563 \\
        SMAPGAN & 290.65 &  \underline{24.918} &  347.38 &  \underline{27.564} \\
        CreativeGAN & 172.74 &  23.204 &  149.34 &  25.683 \\
        LCAG   & \underline{134.05} &  23.876 &  \underline{125.8}  &  25.409 \\
        SCGM-2X & 114.21 &  29.204 &  124.46 &  27.318 \\
        SCGM-4X & \textbf{103.99} & \textbf{29.759} & \textbf{104.63} & \textbf{28.012} \\
        \bottomrule
    \end{tabular}
\end{table}

\subsubsection{Qualitative evaluation}

Visual comparisons (Figure \ref{fig:experiment-qualitative}) reveal significant limitations in existing methods. Due to tile-edge artifacts and inconsistent object granularity, single-scale approaches fail to maintain coherent geographic features, such as fragmented roads and discontinuous vegetation patterns, particularly at small scales (1:4,000 and 1:2,000, in Figure \ref{fig:experiment-qualitative}(b) and (c)). Although multi-scale methods, such as SMAPGAN and LCAG, improve hierarchical representation, their tile results have obvious visual discontinuity problems after map splicing, and the features of the objects they express also have fractures and noise, thus undermining cartographic utility (Figure \ref{fig:experiment-qualitative}(d)). In contrast, SCGM achieves seamless spatial continuity and scale-aware fidelity. For instance, maps at a scale of 1:2,000 preserve intricate urban details, such as building footprints and road networks, while those at 1:8,000 generalize structures without compromising geometric accuracy (Figure \ref{fig:experiment-qualitative}(e)).

Cross-scale comparison of the generation outcomes at different scales reveals that other methods produce similar ground object representations, represent a problem of inconsistency between map synthesis and map scale. By contrast, our method demonstrate superior synthesis features across various scales: generated tile maps closely align with GT (ground truth) in terms of local coherence (e.g., unbroken linear features) and global consistency (e.g., simplification of geographical features), validating its ability to harmonize multi-scale semantics with generative mapping process. These results underscore that our method significantly outperforms the comparison methods in generating multi-scale maps from remote sensing imagery, thereby validating the effectiveness of our approach.

\subsubsection{Map features evaluation}

\begin{table}[ht]
    \caption{Map features evaluation results on the CSCMG dataset.}
    \label{tab:mfp}
    \centering
    \begin{tabular}{lllll}
        \toprule
        Model       &FID↓  &SSIM↑ &PSNR↑ &MFP↑\\
        \midrule
        Pix2pix     &132.62 &0.7174 &23.29	&0.4834\\
        CycleGAN    &109.57	&0.7031	&24.10	&0.5357\\
        SMAPGAN     &178.3  &\textbf{0.8204}	&24.29	&0.3538\\
        ATME        &101.99 &0.7454	&22.65	&0.6600\\
        TSIT        &109.25 &0.7142 &\textbf{24.87}  &0.5910\\
        SCGM        &\textbf{31.60}  &0.7503 &24.79  &\textbf{0.7351}\\
        \bottomrule
    \end{tabular}
\end{table}

To thoroughly evaluate the interpretability and spatial consistency of SCGM, we conducted a map feature assessment using the MFP metric (Section \ref{sec:metrics}) on the CSCMG dataset. Our comparative analysis extended to include representative generative models across two paradigms: GANs, such as Pix2pix, CycleGAN, SMAGAN, and TSIT \citep{jiang2020tsit}, as well as DMs, including ATME and our proposed SCGM. This comprehensive comparison validates the discriminative power of the MFP metric while highlighting the strengths of SCGM. All models were trained on identical datasets and used the same strategies to ensure fairness. Table \ref{tab:mfp} summarizes the quantitative results for 1,000 generated map tiles.

The MFP metric—comprising global feature similarity ($\mathcal{L}_G$) and spatial correlation ($\mathcal{L}_S$)—provides critical insights into semantic alignment and structural fidelity (Section \ref{sec:metrics}). SCGM achieved the highest MFP score of 0.7351, significantly outperforming all baselines. Notably, it surpassed the diffusion-based ATME by 11.4\%, demonstrating superior preservation of cartographic semantics, such as road topology and building distributions. While GAN-based models exhibited pixel-level similarities (e.g., Pix2pix and CycleGAN attained comparable PSNR values of 23.29 and 24.10, respectively), their low MFP scores (0.4834 and 0.5357) reflect inherent limitations in spatial-semantic consistency. This disparity underscores MFP's sensitivity to semantic discrepancies that conventional pixel-wise metrics fail to capture. For instance, SMAGAN achieved competitive SSIM (0.8204) and PSNR (24.29) scores but registered the lowest MFP (0.3538), indicating that it produced structurally plausible yet semantically inconsistent outputs. In contrast, SCGM's balanced excellence across all metrics, including a leading FID score of 31.60 and an SSIM score of 0.7503, validates its comprehensive generation capability.

Traditional metrics further support these findings. The substantial FID gap between SCGM (31.60) and TSIT (109.25) highlights SCGM's alignment with real map distributions. While SCGM's PSNR (24.79) is slightly lower than TSIT's (24.87), this marginal difference highlights that pixel-level fidelity alone cannot guarantee cartographic validity—a limitation addressed by MFP's emphasis on spatial-topological relationships. The metric's quantification of self-attention key correlations in spatial errors provides critical insights into model behavior. Our analysis establishes that SCGM advances the state-of-the-art in multi-scale map generation by coordinating map scale modalities with feature representation. It positions SCGM as a robust solution for high-fidelity map generation, supported by both quantitative rigor and qualitative plausibility.

\subsection{Ablation studies}

In this section, we perform ablation studies on two core components of SCGM: the cascading feature encoder and the scale-modality encoder, as well as on cascade spans (2X/4X), to validate their necessity and optimal configurations. All experiments are conducted on the CSCMG dataset.

\begin{figure*}[htp]
    \centering
    \includegraphics[width=\linewidth]{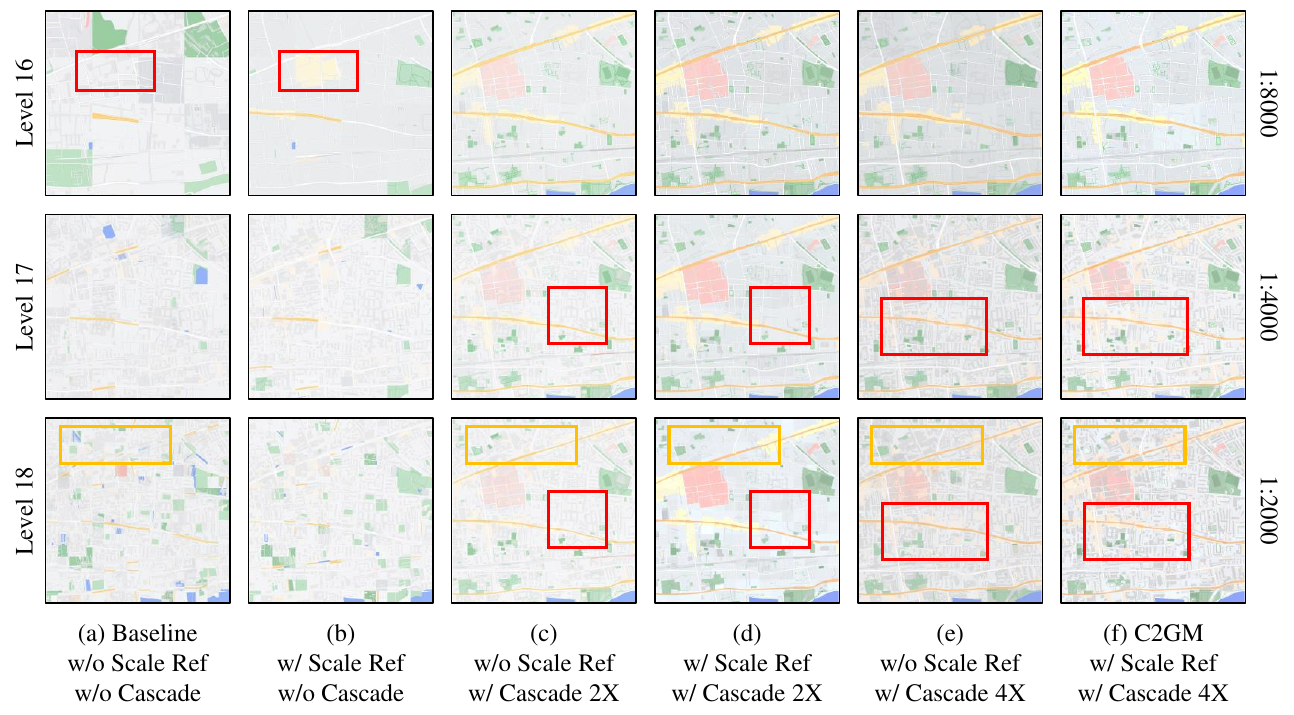}
    \caption{Incremental visual results showcasing seamless, high-quality tile map generation with the addition of cascade referencing (MFEncoder) and scale encoding (ScaleEncoder). The orange box highlights fine-grained map details contributed by cascade references, while the red box denotes enhanced scale representation from scale encoding.}
    \label{fig:experiment-ablation}
\end{figure*}

\begin{table}[t]
    \caption{Ablation results for MFEncoder and ScaleEncoder on CSCMG dataset, where "w/ 2X" and "w/ 4X" indicate conditional generation using 2X or 4X cascade references, respectively.}
    \label{tab:ablation}
    \centering
    \begin{tabular}{lllll}
        \toprule
        +MFEncoder & +ScaleEncoder & PSNR$\uparrow$ & SSIM$\uparrow$ & FID$\downarrow$ \\
        \midrule
        \hspace{8pt}w/o     & \hspace{8pt}w/o   & 19.962 & 0.873 & 52.543 \\
        \hspace{8pt}w/o     & \hspace{8pt}w/    & 20.898 & 0.882 & 44.217 \\
        \hspace{8pt}w/ 2X   & \hspace{8pt}w/o   & 23.096 & 0.934 & 45.512 \\
        \hspace{8pt}w/ 2X   & \hspace{8pt}w/    & 31.069 & 0.941 & 50.052 \\
        \hspace{8pt}w/ 4X   & \hspace{8pt}w/o   & 24.886 & 0.915 & 44.329 \\
        \hspace{8pt}w/ 4X   & \hspace{8pt}w/    & \textbf{31.584} & \textbf{0.935} & \textbf{34.369} \\
        \bottomrule
    \end{tabular}
\end{table}

\subsubsection{Effects of MFEncoder}

Integrating cascade references into the MFEncoder significantly enhances map generation quality, as evidenced by quantitative metrics (Table \ref{tab:ablation}). FID scores for generated maps with 2X and 4X cascade references are 45.512 and 44.329, respectively, while PSNR scores are 23.096 and 24.886. Compared to the scenario without cascade references, these FID scores reflect increases of 7.031 and 8.241, and PSNR scores show improvements of 3.134 and 4.924, respectively. Such results demonstrate that cascade generation significantly enhances the quality of the generated maps. Furthermore, an improved SSIM index indicates enhanced structural similarity of the generated maps, leading to more accurate outputs.

From Figures \ref{fig:experiment-ablation} (a), (c) and (e), it is evident that considering multi-scale information mitigates misjudgments of geographic elements that arise when only current-scale data is analyzed. Visual discontinuities of geographic elements in the original map have been corrected, bringing the generated map closer to the actual representation. Results across different scales illustrate that employing a multi-branch cartographic feature encoder with cascade references allows for integrating single-scale generated maps with information from other scales. This integration enables the mapping model to better recognize objects' comprehensive features across various scales, thus acquiring more extensive cartographic expertise.

Moreover, Figures \ref{fig:experiment-ablation} (d) and (f) demonstrate that comparing the 2X and 4X cascade generation results from the ablation tests reveals superior outcomes for the 4X cascade generation in terms of FID, PSNR, and SSIM metrics. A broader cascade generation span facilitates the model's ability to learn comprehensive map features more effectively, resulting in more accurate maps. The performance of the 2X cascade generation may be limited by its narrower span, which could restrict the diversity of comprehensive map features the generation model can learn. Additionally, relatively similar cascade references may interfere with the terrain features extracted from the images, thereby diminishing the quality enhancement effect of the map cascade generation.

\subsubsection{Effects of ScaleEncoder}

The ScaleEncoder ensures an accurate representation of scale-specific cartographic semantics. In a map with a scale of 1:8000, the ScaleEncoder effectively removes erroneous elements such as an inappropriate yellow background and nonexistent roads. As shown in Table \ref{tab:ablation}, incorporating the ScaleEncoder results in improvements across all metrics compared to the baseline, particularly an 8.3 increase in the FID score, highlighting the significant impact of scale information guidance on accurately representing map features. The CLIP-based scale information encoder translates the textual scale data of the generated map into image features, integrates these with ground features extracted from remote sensing images, and embeds them into the pre-trained generator. This process enables the model to discern comprehensive differences among maps of varying scales, leading to improved generation outcomes at each scale that more accurately reflect the characteristics specific to those scales.

Figure \ref{fig:experiment-ablation} illustrates the results of the ablation experiment. In particular, Figures \ref{fig:experiment-ablation}(a), (c), and (e) display the baseline map denoising results based on remote sensing images, whereas Figures \ref{fig:experiment-ablation}(b), (d), and (f) show the results after integrating scale information. Maps created with scale information demonstrate significant generalization differences at varying scales compared to those generated without it.

Specifically, in Figure \ref{fig:experiment-ablation}(a), the baseline method at a scale of 1:8000 includes superfluous elements like a yellow background and fictitious roads, which are effectively removed after integrating scale information, as seen in Figure \ref{fig:experiment-ablation}(b). Similarly, in Figures \ref{fig:experiment-ablation}(c) and \ref{fig:experiment-ablation}(e), as well as Figures \ref{fig:experiment-ablation}(d) and \ref{fig:experiment-ablation}(f), the baseline method at a scale of 1:2000 incorporates unnecessary buildings and roads, which are also removed after incorporating scale information. Therefore, maps generated by the baseline method exhibit considerable feature discrepancies across different scales, which are effectively corrected by integrating scale information.

This finding indicates that the model can effectively discern differences in map features across scales. Maps generated without scale information exhibit substantial errors, such as incorrect placement of roads and buildings and the presence of non-existent elements. These inaccuracies result from the model’s inability to accurately capture scale-dependent features, leading to incoherent results. Incorporating scale information rectifies these issues, producing a more accurate map at a given scale.

\section{Discussion}

\begin{figure*}[htp]
    \centering
    \includegraphics[width=\linewidth]{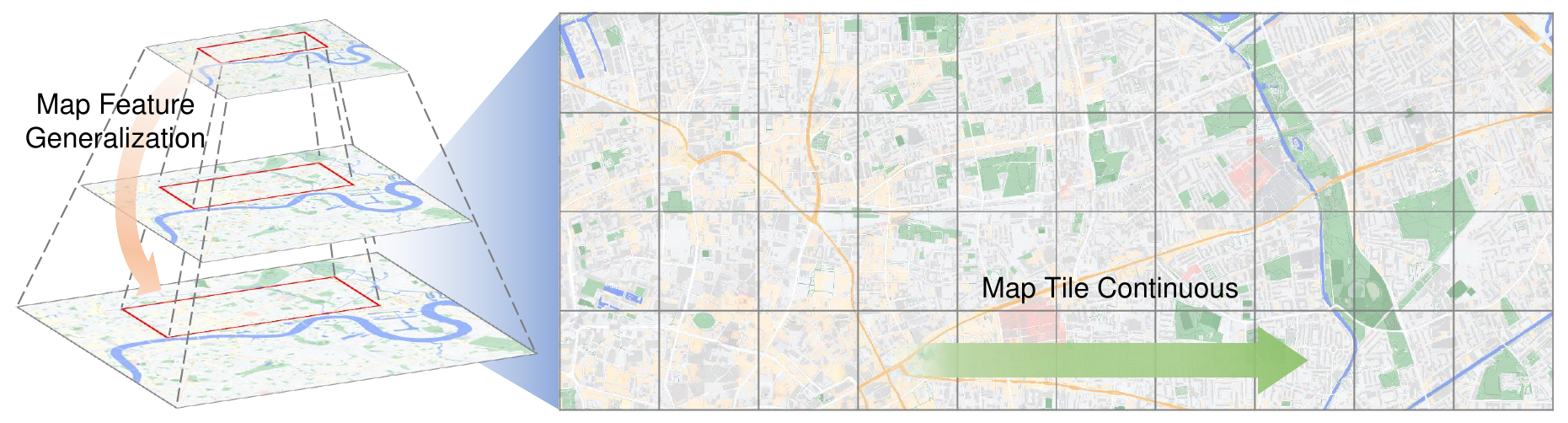}
    \caption{Seamless tile map generation: SCGM demonstrates powerful capabilities on generating seamless, multi-scale, large-format, and virtually limitless tile maps.}
    \label{fig:discussion-map}
\end{figure*}

\subsection{Analysis of multi-scale cascade generation}

The integration of cascade references enhances multi-scale feature generalization. By leveraging small-scale tiles (e.g., 1:15,000 scale) as priors, SCGM generates higher-scale outputs (e.g., 1:8000 to 1:2,000) with enriched geographic semantics (Figure \ref{fig:experiment-ablation}(f)). This cascaded refinement captures scale-specific feature hierarchies—such as road network granularity and vegetation patterns—more effectively than fixed-layer constraints \citep{li2020mapgan,fu2022translation}. The framework’s ability to simulate cartographic generalization processes enables the synthesis of realistic, seamless tile maps across diverse scales (Figure \ref{fig:discussion-map}), supporting applications like disaster response and urban planning where accuracy and detail adaptability are paramount.

Furthermore, the cascading generation strategy in SCGM addresses critical challenges in large-format multi-scale map synthesis. Unlike conventional tile stitching methods \citep{liu2021cscgan,fu2023levelaware}, which introduce visual discontinuities at feature boundaries (e.g., roads, buildings), SCGM enforces spatial coherence by conditioning high-level tile generation on lower-scale outputs (Figure \ref{fig:discussion-map}). This approach mitigates edge artifacts common in independent tile generation \citep{chen2021smapgan}, ensuring stylistic consistency and geographic continuity across scales. For instance, urban maps at 1:2,000 scale retain intricate details (e.g., building footprints), while 1:8,000 scale maps exhibit generalized structures, mirroring cartographic conventions (Figure \ref{fig:experiment-qualitative}(k)).

\subsection{Limitations and environmental adaptability}

While SCGM excels in urban mapping (scales 1:8,000 to 1:2,000) by handling artificial features (e.g., buildings, roads) through aggregation and simplification, its performance declines in natural landscapes and small-scale mapping (<1:35,000). For natural features (e.g., forests, terrain), scale transitions rely on gradual geometric smoothing rather than structural reorganization, limiting cascaded learning efficacy due to inherent self-similarity. Tile-based generation assumes localized spatial coherence, which suits urban layouts but fails for expansive terrains with long-range dependencies. At small scales (<1:35,000), geometric abstraction converts features into symbolic representations, challenging pixel-level cascade learning. Although CLIP-based scale encoding mitigates global inconsistencies, the absence of terrain-aware attention mechanisms hinders geomorphological fidelity. Future work should focus on enhancing natural scene modeling while preserving urban mapping strengths.

\section{Conclusion}

This paper introduces SCGM, an innovative foundational framework for generative mapping designed to create multi-scale tile map from remote sensing images. The primary contributions of this study consist of the following: First, a CLIP-guided scale-modality encoding mechanism that dynamically aligns cartographic generalization rules (e.g., resolution, legends) with geographic semantics. Second, a cascaded generation strategy that enforces cross-tile consistency by conditioning larger-scale outputs on smaller-scale priors, eliminating tile discontinuities. Third, a dual-branch adaptation network that hierarchically fuses multi-scale features while preserving geographic fidelity. When evaluated on the CSCMG dataset, SCGM achieves state-of-the-art performance in multi-scale tile map generation. It ensures seamless continuity and incorporates scale-aware cartographic features, as measured by the map feature perception metric. The framework’s capability to simulate cartographic workflows opens new avenues for real-time applications in emergency mapping and urban planning.


\section*{Data availability}
The SCGM framework and CSCMG dataset for our key contributions will be open sourced in the code repository: \href{https://github.com/Magician-MO/SCGM}{https://github.com/Magician-MO/SCGM}.




\bibliographystyle{cas-model2-names}

\bibliography{refs}

\end{document}